\documentclass[prl,twocolumn,showpacs,floatfix,amsmath,amssymb,superscriptaddress]{revtex4-1}
\usepackage{amsfonts}
\usepackage{stmaryrd}
\usepackage{bbm}
\usepackage{mathrsfs}
\usepackage{tipa}
\usepackage{amssymb}
\usepackage{txfonts}
\usepackage{graphicx}
\usepackage{dcolumn}
\usepackage{epstopdf}
\usepackage[colorlinks,linkcolor=blue,urlcolor=blue,citecolor=blue]{hyperref}
\usepackage{multirow}
\usepackage{subfigure}
\usepackage{url}
\usepackage{setspace}

\begin{document}

\title{Unconventional charge density wave and photoinduced lattice symmetry change in Kagome Metal CsV$_3$Sb$_5$ probed by time-resolved spectroscopy}
\author{Z. X. Wang}
\affiliation{International Center for Quantum Materials, School of Physics, Peking University, Beijing 100871, China}

\author{Q. Wu}
\affiliation{International Center for Quantum Materials, School of Physics, Peking University, Beijing 100871, China}

\author{Q. W. Yin}
\affiliation{Department of Physics and Beijing Key Laboratory of Opto-electronic Functional Materials and Micro-nano Devices, Renmin University of China, Beijing 100872, China}

\author{Z. J. Tu}
\affiliation{Department of Physics and Beijing Key Laboratory of Opto-electronic Functional Materials and Micro-nano Devices, Renmin University of China, Beijing 100872, China}

\author{C. S. Gong}
\affiliation{Department of Physics and Beijing Key Laboratory of Opto-electronic Functional Materials and Micro-nano Devices, Renmin University of China, Beijing 100872, China}

\author{T. Lin}
\affiliation{International Center for Quantum Materials, School of Physics, Peking University, Beijing 100871, China}

\author{Q. M. Liu}
\affiliation{International Center for Quantum Materials, School of Physics, Peking University, Beijing 100871, China}

\author{L. Y. Shi}
\affiliation{International Center for Quantum Materials, School of Physics, Peking University, Beijing 100871, China}

\author{S. J. Zhang}
\affiliation{International Center for Quantum Materials, School of Physics, Peking University, Beijing 100871, China}

\author{D. Wu}
\affiliation{International Center for Quantum Materials, School of Physics, Peking University, Beijing 100871, China}

\author{H. C. Lei}
\affiliation{Department of Physics and Beijing Key Laboratory of Opto-electronic Functional Materials and Micro-nano Devices, Renmin University of China, Beijing 100872, China}

\author{T. Dong}
\affiliation{International Center for Quantum Materials, School of Physics, Peking University, Beijing 100871, China}

\author{N. L. Wang}
\affiliation{International Center for Quantum Materials, School of Physics, Peking University, Beijing 100871, China}
\affiliation{Beijing Academy of Quantum Information Sciences, Beijing 100913, China}

\begin{abstract}

Recently, kagome lattice metal AV$_3$Sb$_5$ (A = K, Rb, Cs) family has received wide attention due to its presence of superconductivity, charge density wave (CDW) and peculiar properties from topological nontrivial electronic structure. With time-resolved pump-probe spectroscopy, we show that the excited quasiparticle relaxation dynamics can be explained by formation of energy gap below the phase transition being similar to a usual second-order CDW condensate, by contrast, the structure change is predominantly first order phase transition. Furthermore, no CDW amplitude mode is identified in the ordered phase. The results suggest that the CDW order is very different from the traditional CDW condensate. We also find that weak pump pulse can non-thermally melt the CDW order and drive the sample into its high temperature phase, revealing the fact that the difference in lattice potential between those phases is small.
\end{abstract}


\maketitle

Control over physical properties or electronic phases in quantum materials via external tuning parameters is a central topic in condensed matter physics. Traditional ways of tuning external parameters are to change the temperature, pressure, electric/magnetic field, or doping level of the material systems. In the last two decades, photo-excitation has emerged as a new way to probe and manipulate the properties of quantum materials, enabling an ultra-fast control over the material properties or quantum phases \cite{Cavalleri2001,Tomimoto2003,Okamoto2004,Takubo2005,Matsubara2007,Beyer2009,Wall2012,Stojchevska177,Giannetti2016,Zhang2019,Sie2019,Nova2019,Li2019,McLeod2020,Kogar2020,Liu2021}. The ultrashort laser pulses can create a far from equilibrium distribution of the energy among the different degrees of freedom, triggering the formation of thermal or non-thermal metastable or even stable phases. Among different systems, correlated electronic systems with rich phase diagrams have been the most explored materials for practical manipulation, since the complex interactions or competition among different degrees of freedom make them very susceptible to external perturbation or stimuli \cite{Giannetti2016}.

Recently, a new family of kagome metals AV$_3$Sb$_5$ (A = Cs, Rb, K), composed of vanadium layers, antimony layers and alkali ions sandwiched between the two layers, has attracted tremendous attention in the community \cite{Ortiz2019,Ortiz2020,Ortiz2021,Yin2021}. These materials undergo a charge density wave (CDW) phase transition at $T_{\rm{CDW}}=80-100$ K with a 2$\times$2$\times$2 superlattice formation and a superconducting transition at $T_{\rm{c}}=0.9-3.5$ K. High-pressure measurements revealed the competition between the charge order and superconductivity by observing a double-dome superconductivity\cite{Zhao2021,Chen2021,Du2021,Chen2021a,Zhang2021}. More intriguingly, the topological nontrivial electronic structure was found in AV$_3$Sb$_5$ family (even in CDW state) \cite{Ortiz2019,Ortiz2020,Ortiz2021,Yin2021,Wang2020,Tan2021,Li2021,Zhao2021a,Chen2021b,Ni2021,Jiang2020}. A giant anomalous Hall effect \cite{Yang2020,Yu2021} was observed even in the absence of magnetic ordering \cite{Ortiz2020,Kenney2020}. Furthermore, transport and scanning tunnelling microscopy (STM) measurements revealed that the 6-fold rotation $C_6$ symmetry is further broken and reduced to $C_2$ symmetry at low temperature below 60 K \cite{Xiang2021,Zhao2021,Chen2021b,Li,Ratcliff2021}. A recent coherent phonon spectroscopy measurement \cite{Ratcliff2021} revealed appearance of new phonon modes below 94 K and 60 K, respectively, implying structural phase transitions. Thus, the AV$_3$Sb$_5$ series provide a new opportunity to study the competition between different phases and to understand the unconventional correlated physics emerging from the itinerant kagome lattice electrons.

In this work, we perform ultrafast pump-probe reflectivity experiment on CsV$_3$Sb$_5$, aiming to investigate the CDW order across the phase transitions and possible photoexcitation control of different orders in the compound. Our measurements show that, while the quasiparticle relaxation dynamics upon weak pumping can be described by formation of energy gap below the phase transition, the structure change is characterized by an abrupt change in the number of coherent phonon modes without showing clear softening at the CDW phase transition. Furthermore, no CDW amplitude mode can be identified in the ordered phase. Those results suggest that the CDW order is predominantly first order phase transition, and is very different from the traditional CDW condensate. More intriguingly, we show that even small pumping fluence can non-thermally melt the $C_2$ ground state order and $C_6$ CDW order successionally and drive the compound into high temperature phase. We shall discuss the implication of the results.

Single crystals of CsV$_3$Sb$_5$ were grown from Cs ingot (purity 99.9\%), V powder (purity 99.9\%) and Sb grains (purity 99.999\%) using the self-flux method, similar to the growth of RbV$_3$Sb$_5$ \cite{Yin2021}. We used an amplified Ti:sapphire laser system with 800 nm wavelength and 35 fs pulse duration operating at 1 kHz repetition frequency as the light source for pump-probe measurement. The fluence of probe beam is set below 3 $\mu$J cm$^{-2}$, weaker than that of pump beam. To reduce the noise caused by stray light, the pump and probe pulses were set to be cross polarized and an extra polarizer was mounted just before the detector.

\begin{figure}[htbp]
	\centering
	\includegraphics[width=9cm]{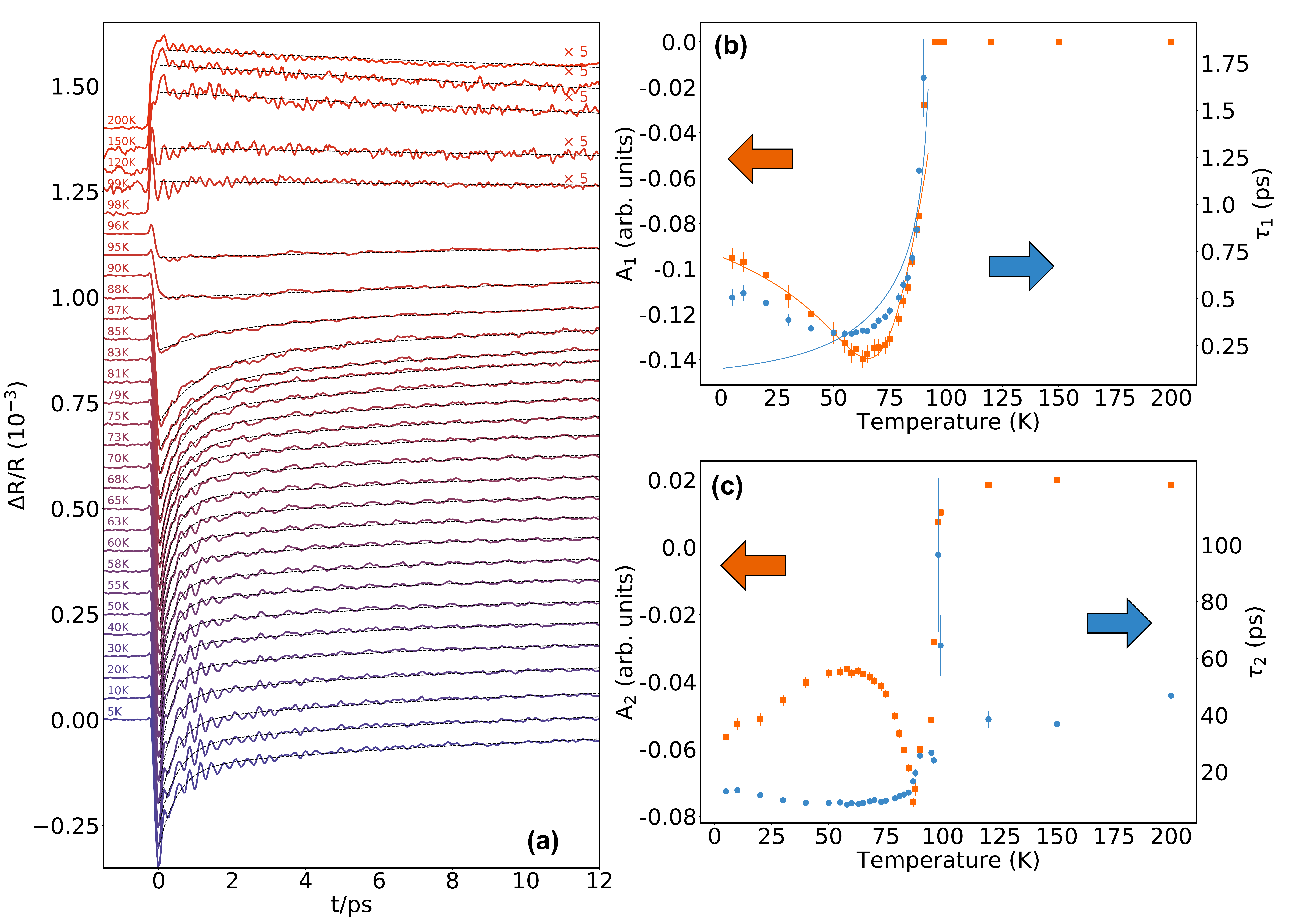}\\
	\caption{\textbf{Temperature-dependent dynamics for CsV$_3$Sb$_5$.} (a) $\Delta R/R$ in the temperature range of 5-200 K. With increasing temperature, the absolute value of $\Delta R/R$ increases first, then decreases and finally it changes the sign at $T \approx 98 $ K (above $T_{\rm{CDW}}=94$ K). The lines in high temperature phase are scaled by a five-fold factor. The dash lines are the double-exponential fitting curve: $\Delta R/R(t) = A_1 \rm{e}^{-t/\tau_1}+A_2 \rm{e}^{-t/\tau_2}$. (b) (c) A fast ($\tau_1\sim0.5$ ps) and a slow ($\tau_2\sim10$ ps) relaxation process, respectively. The amplitude $A_1$, $A_2$ (orange squares) and decay time $\tau_1$, $\tau_2$ (blue circles) of the reflectivity transients extracted from fits to the double-exponential. We notice the amplitude of the fast relaxation process $A_1$ becomes 0 in high temperature phase, which means the fast relaxation process is absent in high temperature phase and the double-exponential function degrades into single-exponential function. Error bars represent the standard deviation of the fit. The orange and blue lines are the RT model fits.}\label{Fig:1}
\end{figure}

\begin{figure}[htbp]
	\centering
	\includegraphics[width=8cm]{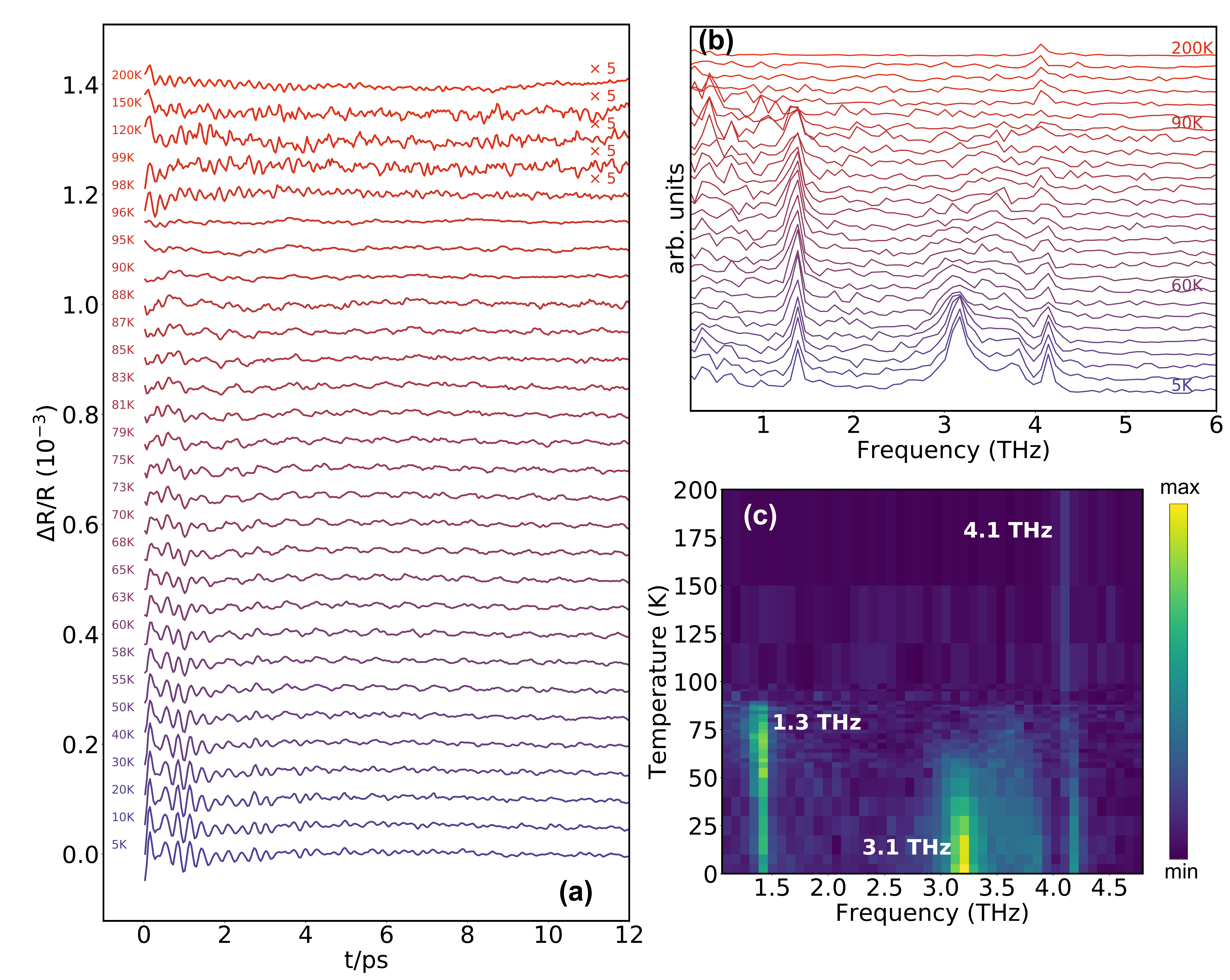}\\
	\caption{\textbf{Temperature-dependent coherent phonon spectroscopy for CsV$_3$Sb$_5$.}
	(a) Coherent phonon oscillation in time-domain, where the decay background is subtracted.
	(b) The Fast Fourier transformation of the data in (a). (c) Temperature dependence waterfall map extracted from (b). The 4.1 THz coherent phonon is present at all temperature through phase change. The 1.3 THz phonon can be only detected below $T_{\rm{CDW}}$, while the 3.1 THz phonon disappears at the temperature above $T^{*}=30\sim60$ K.
	}\label{Fig:2}
\end{figure}

Figure \ref{Fig:1} presents the photoinduced reflectivity change of CsV$_3$Sb$_5$ as a function
of time delay at different temperatures under rather weak pumping excitation $\sim$5 $\mu $J cm$^{-2}$. Very prominently, $\Delta R/R$ changes sign, from positive to negative when temperature decreases across the CDW transition temperature $T_{\rm{CDW}}$. At low temperature, the pump pulse induces an abrupt drop in reflectivity followed by a fast decay within a few picoseconds and a slower recovery process in the order of 10 ps, while there is only a slow decay dynamics at high temperature. We use double-exponential function to fit the decay process: $\Delta R/R = A_1 \rm{e}^{-t/\tau_1}+A_2 \rm{e}^{-t/\tau_2}$, where $A_1 \rm{e}^{-t/\tau_1}$ is the fast decay process and $A_2 \rm{e}^{-t/\tau_2}$ presents the slower recovery process, respectively. The best fitting results are shown as dashed lines in Fig.\ref{Fig:1} (a). We note the fast decay process becomes invisible above 95 K, which indicates the strong correlation with CDW energy gap. Fitting parameters of the double-exponential function are shown in Fig. \ref{Fig:1} (b) and (c). For the fast decay dynamics, the decay lifetime $\tau_1$ diverges and the absolute value of amplitude $A_1$ drops to zero precipitously, as the temperature rising close to $T_{\rm{CDW}}$. These behaviors indicate that an energy gap is closing toward critical temperature, which can be roughly described by Rothwarf-Taylor (R-T) model, showing that the opening of a CDW energy gap would significantly impede the relaxation of the photoinduced quasiparticle \cite{Kabanov1999}. Similar to the situation of superconductivity from the condensate of electron-electron Cooper pairs, CDW is a condensate from electron-hole (e-h) pairs. The pump pulse serves as an excitation to break e-h pairs across energy gap of CDW condensate, the recombination of e-h pairs is accompanied by the emission of phonons with energy higher than the energy gap $\Delta(T)$. $\tau_1$ is the characteristic time for this recombination, since these excited phonons will in turn break additional e-h pairs, decay time $\tau_1$ is ultimately determined by the time required for these phonons anharmonically decaying into phonons with the energy less than $\Delta$. The amplitude $A$ represents the population of excited quasiparticles by pump pulse. The change of relaxation time and the amplitude of the photoinduced reflectivity signal near the transition temperature in R-T model is given by following equations \cite{Kabanov1999,PhysRevB.101.205112}:
\begin{equation}
\begin{aligned}
    \tau(T) &\propto \frac{\ln{\left[ g+\textrm{e}^{-\Delta(T)/k_B T} \right]}}{\Delta(T)^2}\\
    A(T) &\propto \frac{\Phi/(\Delta(T)+k_B T/2)}{1+\Gamma \sqrt{2k_B T /\pi \Delta(T)}\textrm{e}^{-\Delta(T)/k_B T}}
\end{aligned}
\end{equation}
where a BCS-type energy band is assumed here as $\Delta(T)=\Delta_0\sqrt{1-T/T_{\rm{CDW}}}$ with $T_{\rm{CDW}}=94$ K and the CDW gap at 0 K $\Delta_0=40$ meV \cite{Wang2021}. Qualitatively, the equations can reproduce the measurement results. The best fits yield the phenomenological fitting parameter $g=0.18$ and $\Gamma=17$ as shown in Fig. \ref{Fig:2} (b). Although there appears an increase of the decay time $\tau_1$ below $T^*=60$ K, no significant change can be identified from the relaxation process when the symmetry is further broken from  $C_6$ to $C_2$.

\begin{figure*}[htbp]
 \centering
 \centering\includegraphics[width=14cm]{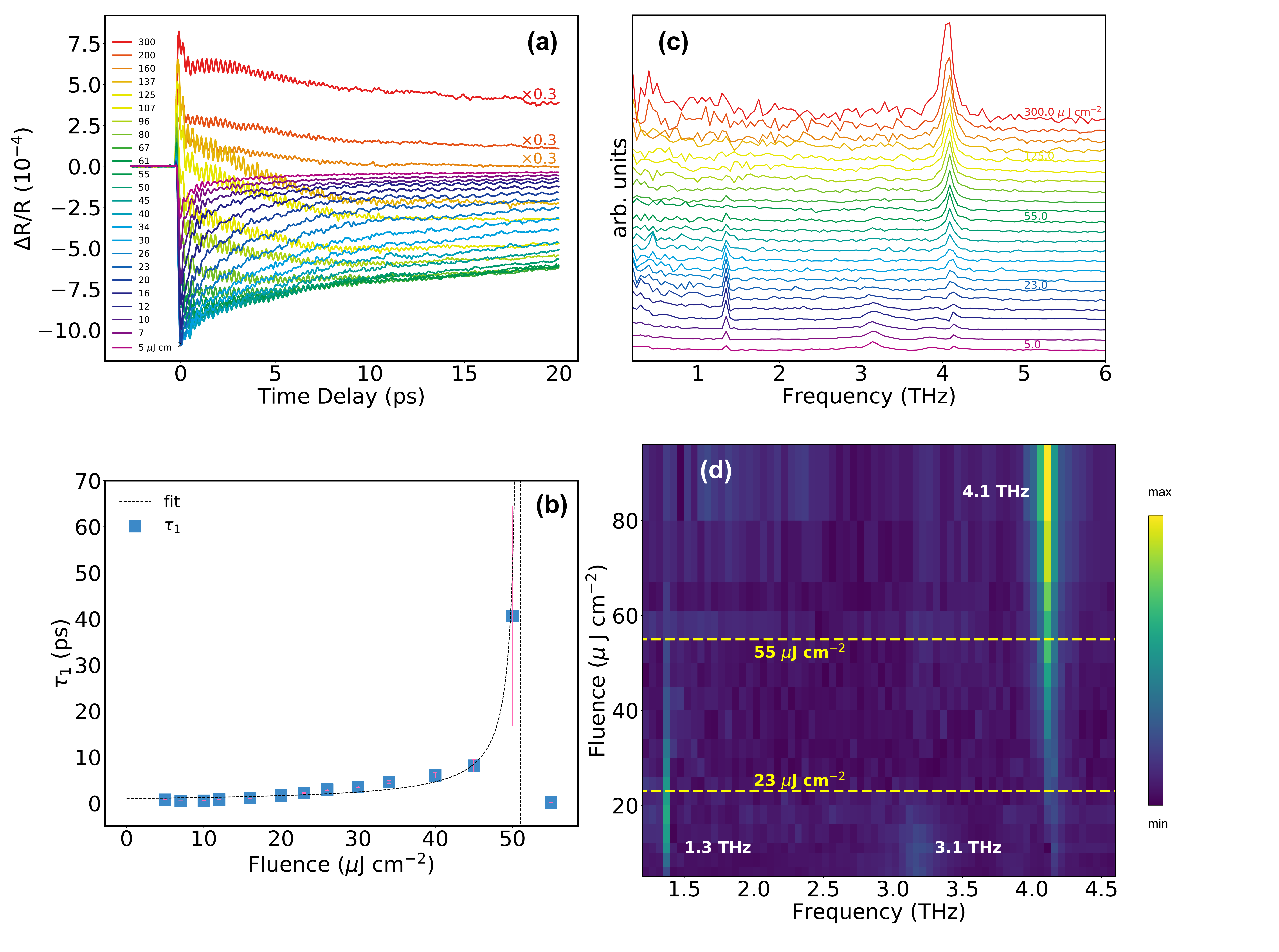}\\
 \caption{
\textbf{Fluence-dependent pump probe reflectivity signal of CsV$_3$Sb$_5$.} (a) Photo-induced reflectivity $\Delta R/R$ as a function of time delay versus pump fluence in range from 5 to 300 $\mu$J cm$^{-2}$ of CsV$_3$Sb$_5$ at 20 K. The data of the three highest fluence is scaled by a factor of 0.3. (b) Fluence dependence of decay time $\tau$ of the reflectivity transients extracted from fits to the double exponential
decay function. (c) The Fast Fourier transformation of the oscillation parts by subtracting the decay background of $\Delta R/R$ in (a). Three resonance peaks at 1.3, 3.1 and 4.1 THz can be seen at the lowest fluence 5 $\mu$J cm$^{-2}$. (d) Two-dimensional waterfall map of these three resonance peaks. 3.1 THz phonon can be survived below the fluence of 23 $\mu$J cm$^{-2}$, the phonon at 1.3 THz is killed by 55 $\mu$J cm$^{-2}$ pulses, while the 4.1 THz phonon is coherently enhanced even under an intense fluence higher than 300 $\mu$J cm$^{-2}$.} \label{Fig:4}
\end{figure*}

After subtracting the dynamics background, we obtained the coherent phonon oscillations, as shown in Fig. \ref{Fig:2} (a). The fast Fourier transform result is presented in Fig. \ref{Fig:2} (b). A two dimensional intensity map as a function of temperature and frequency is displayed in Fig. \ref{Fig:2} (c). The coherent phonon spectroscopy is similar to previous study on CsV$_3$Sb$_5$ \cite{Ratcliff2021}. Above $T_{\rm{CDW}}=94$ K, there is only one peak at 4.1 THz, which is in good agreement with the density functional theory (DFT) calculations for the $A_{1g}$ mode of phonon spectra \cite{Li2021,Ratcliff2021}. Below $T_{\rm{CDW}}$, a new mode at 1.3 THz peak abruptly appears, yielding evidence for the CDW structural modulation. Below 60 K, a weak and broad feature emerges near 3.1 THz, which evolves into a sharp and strong peak below 30 K. This mode was suggested to be linked to the uniaxial order observed by STM experiments which breaks the $C_6$ rotational symmetry \cite{Ratcliff2021}. It deserves to remark that, for conventional CDW condensate, the CDW amplitude mode usually appears as the strongest oscillation in the pump-probe measurement. As the temperature increases to the transition temperature $T_{\rm{CDW}}$, the mode frequency softens and behaves like an order parameter \cite{PhysRevLett.118.107402}. However, no such CDW amplitude mode could be identified from the above spectra, suggesting that the CDW order in CsV$_3$Sb$_5$ is very different from the conventional CDW condensate. The abrupt change of the coherent mode near 1.3 THz further indicates that the phase transition is predominantly first-order phase transition. On the other hand, the symmetry change from $C_6$ to $C_2$ is a crossover behavior. A short range of $C_2$ nematic order, corresponding the broad feature, is seen roughly below 60 K which gradually evolves into a static $C_2$ order below 30 K.

The above analysis reveals a rather peculiar situation: while the excited electron relaxation dynamics upon weak pumping can be described by formation of energy gap below the phase transition being similar to a usual second-order CDW condesate, the structure change is primarily first order phase transition, being characterized by an abrupt change in the number of coherent phonon modes without showing clear softening at the CDW phase transition. Additionally, no CDW amplitude mode is present in the ordered phase. Those results suggest that the CDW order is rather unconventional. We notice that nuclear magnetic resonance (NMR) measurements on CsV$_3$Sb$_5$ indicated similar situation \cite{Song}. A first-order structural phase transition associated with orbital ordering is seen as the sudden splitting of orbital shift in $^{51}$V NMR spectrum at T$_{CDW}$, by contrast, a typical second order transition behavior is seen in the quadrupole splitting which appears gradually below T$_{CDW}$. The NMR measurement also suggests that the CDW order is a secondary electronic order. The seemingly week coupling between electron and lattice degree of freedoms needs to be further explored.

In addition to varying temperature, we also performed measurement of the photoinduced reflectivity change at different pump fluences in CDW phase. The measurement results at 20 K are shown in Fig.\ref{Fig:4}. We found that a surprisingly small pump fluence can melt the $C_2$ nematic and then the $C_6$ CDW order. As shown in Fig.\ref{Fig:4} (a), at 5 $\mu $J cm$^{-2}$ fluence, the fast decay process shows a lifetime of hundreds of femtoseconds. With pump fluence incresing, the absolute value of negative $\Delta R/R$ further grows, and reaches the maximum value at around $F^{\rm{melt}}\sim$55 $\mu $ J cm$^{-2}$. Then it turns back and goes into the positive direction, passes through zero, and gradually grows larger. This behavior is very similar to the warming process of temperature dependent measurement. From the two exponential decay approach, we can extract the relaxation time of fast decay dynamics, as shown in Fig.\ref{Fig:4} (b). As the fluence increases to $F^{\rm{melt}}$, the relaxation time sharply increases. Similar to the above analysis for temperature variation, we assume that an CDW energy gap is present up to the melting fluence as $\Delta=\Delta_0(1-F/F^{\rm{melt}})$, and the decay time is inversely proportional to the energy gap near the critical melting fluence $\tau \propto 1/\Delta$. Indeed, the fast decay time can be reasonably reproduced by the relation as shown in Fig.\ref{Fig:4} (b). The fluence dependent relaxation dynamics suggests that a pump flence as small as 55 $\mu $J cm$^{-2}$ can melt the CDW order in CsV$_3$Sb$_5$.

The coherent phonon spectroscopy provides more direct evidence for ultrafast optical melting of those orders in CsV$_3$Sb$_5$. We performed Fourier transformation of the pump-probe measurement after subtracting the relaxation background and obtained the coherent phonon spectra for different fluences, as displayed in Fig.\ref{Fig:4} (c). The intensity plot as function of both frequency and pump fluence is shown in Fig.\ref{Fig:4} (d). At small fluence, for example, at 5 $\mu $J cm$^{-2}$, three resonance peaks centered at 1.3, 3.1 and 4.1 THz can be observed. As the pump fluence rises to 23 $\mu $J cm$^{-2}$, the peak at 3.1 THz being associated with $C_2$ symmetry\cite{Ratcliff2021} disappears. As the fluence increases beyond 55 $\mu $J cm$^{-2}$, the phonon peak at 1.3 THz, which is related to CDW order \cite{Ratcliff2021}, is not visible. The critical pump fluence is in agreement with the analysis form the relaxation dynamics. The 4.1 THz resonance is present in all fluences. The strength of this mode is further enhanced up to the highest used fluence 300 $\mu $J cm$^{-2}$. The behavior is expected normally for the coherent phonon generation in pump-probe experiment.

We emphasize that the melting of the low temperature orders cannot be attributed to trivial thermal effect owning to the very small pump fluence. The temperature rise $\Delta T$ can be phenomenologically estimated by energy conservation law $S \delta_0 \rho /M \int_{T_0}^{T_0 + \Delta T} C_p(T)\textrm{d}T=(1-R)FS$, where $S$ is the excitation area, the mass density $\rho=5.2$ g/cm$^3$, the molar mass $M=894.5$ g/mol, the penetration depth $\delta_0 \approx 64$ nm and the reflectivity $R=0.42$ at 800 nm for CsV$_3$Sb$_5$ from our own reflectance measurement by Fourier transform infrared spectrometer. With the initial temperature $T_0=20$ K and the temperature-dependent thermal capacity $C_p(T)$ is extracted from previous data \cite{Ortiz2020}, the values of $\Delta T$ are calculated to be approximately 10 and 19 K at the fluence of 23 and 55 $\mu$J cm$^{-2}$, respectively. From this estimation, the possibility of thermal effect can be unambiguously ruled out. 

The nonthermal melting at such small pump fluence is the key finding in this work. Although photoinduced lattice symmetry change or phase transitions have been observed in many different systems, the threshold fluences for completely changing the lattice potentials are usually much higher than the present measurement. For example, VO$_2$ undergoes a first order structural phase transition on cooling from a rutile R-phase to a monoclinic M$_1$-phase at 343 K. Photoexcitation at room temperature can induce the lattice symmetry change from the M$_1$-phase to rutile R-phase based on similar coherent phonon measurement \cite{Wall2012}. However, the required pump threshold fluence is $\sim$7 mJ cm$^{-2}$ for 800 nm pump-probe measurement with similar pulse duration $<$40 fs. Our measurement suggests that the lattice potential difference between those different phases are relatively small. The weak pump pulses can excite a sufficient number of electrons and their perturbation to the lattice potential is large enough to modify the symmetry. As a result, it drives the structural phase transitions non-thermally. We noticed that a recent X-ray scattering measurement \cite{Li2021} on CsV$_3$Sb$_5$ revealed that, while the CDW is long-range ordered, the integrated CDW superlattice peak intensity that is proportional to the CDW order parameter is extremely small. Comparing with fundamental Bragg peaks, the CDW peak intensity is 3$\sim$5 orders of magnitude weaker, demonstrating small lattice distortions. This observation appears to be correlated to our result.

To summarize, laser pulses serve as a tool not only to probe the dynamic behavior but also to drive phase transitions in kagome metal compound CsV$_3$Sb$_5$. Our measurement reveals a peculiar CDW phase transition, \emph{i.e.} the quasiparticle relaxation dynamics can be explained by formation of energy gap below the phase transition being similar to a usual second-order CDW condensate, the structure change is predominantly first order phase transition. Furthermore, no CDW amplitude mode can be identified in the ordered phase. We also show that even small pumping fluence can non-thermally melt the $C_2$ ground state order and then $C_6$ CDW order seccessionally and drive the compound into high temperature phase, suggesting that the lattice potential difference between those different phases is relatively small.

\begin{center}
\small{\textbf{ACKNOWLEDGMENTS}}
\end{center}
This work was supported by National Natural Science Foundation of China (No. 11888101, 11822412 and 11774423), the National Key Research and Development Program of China (No. 2017YFA0302904, 2018YFE0202600) and Beijing Natural Science Foundation (Grant No. Z200005)

\bibliography{CsVSb-1}

\end{document}